\documentclass[ALICE,manyauthors]{cernphprep}
\usepackage[comma,square,numbers,sort&compress]{natbib}
\usepackage{hyperref}
\usepackage{lineno}
\usepackage{xspace}
\usepackage{xcolor}
\usepackage[T1]{fontenc}
\usepackage{orcidlink}

\usepackage[separate-uncertainty=true,locale=US,table-align-uncertainty]{siunitx}
\DeclareSIUnit[quantity-product = ]\percent{\char`\%}
\begin{document}
%

\newcommand{\pp}           {pp\xspace}
\newcommand{\OO}           {OO\xspace}
\newcommand{\pO}           {pO\xspace}
\newcommand{\ppbar}        {\mbox{$\mathrm {p\overline{p}}$}\xspace}
\newcommand{\XeXe}         {\mbox{Xe--Xe}\xspace}
\newcommand{\PbPb}         {\mbox{Pb--Pb}\xspace}
\newcommand{\pPb}          {\mbox{p--Pb}\xspace}
\newcommand{\AuAu}         {\mbox{Au--Au}\xspace}
\newcommand{\dAu}          {\mbox{d--Au}\xspace}

\newcommand{\snn}          {\ensuremath{\sqrt{s_{\text{\tiny NN}}}}\xspace}
\newcommand{\asQ}          {\ensuremath{\alpha_{\mathrm{S}}(Q^2)}\xspace}
\newcommand{\pt}           {\ensuremath{p_{\rm T}}\xspace}
\newcommand{\meanpt}       {$\langle p_{\mathrm{T}}\rangle$\xspace}
\newcommand{\ycms}         {\ensuremath{y_{\rm CMS}}\xspace}
\newcommand{\ylab}         {\ensuremath{y_{\rm lab}}\xspace}
\newcommand{\etarange}[1]  {\mbox{$\left | \eta \right |~<~#1$}}
\newcommand{\yrange}[1]    {\mbox{$\left | y \right |~<~#1$}}
\newcommand{\dndy}         {\ensuremath{\mathrm{d}N_\mathrm{ch}/\mathrm{d}y}\xspace}
\newcommand{\dndeta}       {\ensuremath{\mathrm{d}N_\mathrm{ch}/\mathrm{d}\eta}\xspace}
\newcommand{\avdndeta}     {\ensuremath{\langle\dndeta\rangle}\xspace}
\newcommand{\dNdy}         {\ensuremath{\mathrm{d}N_\mathrm{ch}/\mathrm{d}y}\xspace}
\newcommand{\Npart}        {\ensuremath{N_\mathrm{part}}\xspace}
\newcommand{\Ncoll}        {\ensuremath{N_\mathrm{coll}}\xspace}
\newcommand{\dEdx}         {\ensuremath{\textrm{d}E/\textrm{d}x}\xspace}
\newcommand{\RpPb}         {\ensuremath{R_{\rm pPb}}\xspace}
\newcommand{\XSec}         {cross section\xspace}
\newcommand{\XSecs}        {cross sections\xspace}
\newcommand{\RpA}          {\ensuremath{R_{\rm pA}}\xspace}
\newcommand{\ROO}          {\ensuremath{R_{\rm OO}}\xspace}
\newcommand{\RpO}          {\ensuremath{R_{\rm pO}}\xspace}
\newcommand{\ROOpOpO}          {\ensuremath{R_{\rm OO}/R_{\rm pO}^2}\xspace}
\newcommand{\RdAu}          {\ensuremath{R_{\rm dAu}}\xspace}
\newcommand{\RAA}          {\ensuremath{R_{\rm AA}}\xspace}
\newcommand{\RAB}          {\ensuremath{R_{\rm AB}}\xspace}
\newcommand{\RPbPb}        {\ensuremath{R_{\rm PbPb}}\xspace}
\newcommand{\otp}          {\ensuremath{\omega/\pi^0} ratio\xspace}

\newcommand{\nineH}        {$\sqrt{s}~=~0.9$~Te\kern-.1emV\xspace}
\newcommand{\seven}        {$\sqrt{s}~=~7$~Te\kern-.1emV\xspace}
\newcommand{\eight}        {$\sqrt{s}~=~8$~Te\kern-.1emV\xspace}
\newcommand{\thirteen}     {$\sqrt{s}~=~13$~Te\kern-.1emV\xspace}
\newcommand{\thirteensix}  {$\sqrt{s}~=~13.6$~Te\kern-.1emV\xspace}
\newcommand{\twoH}         {$\sqrt{s}~=~0.2$~Te\kern-.1emV\xspace}
\newcommand{\twosevensix}  {$\sqrt{s}~=~2.76$~Te\kern-.1emV\xspace}
\newcommand{\ninesixtwo}   {$\sqrt{s}~=~9.62$~Te\kern-.1emV\xspace}
\newcommand{\five}         {$\sqrt{s}~=~5.02$~Te\kern-.1emV\xspace}
\newcommand{\twosevensixnn}{$\sqrt{s_{\mathrm{NN}}}~=~2.76$~Te\kern-.1emV\xspace}
\newcommand{\fivenn}       {$\sqrt{s_{\text{\tiny NN}}}=5.02$~Te\kern-.1emV\xspace}
\newcommand{\fivethreesix}{$\sqrt{s}~=~5.36$~Te\kern-.1emV\xspace}
\newcommand{\fivethirteen}{$\sqrt{s}~=~5.36$\xspace\text{ and} $13.6$~Te\kern-.1emV\xspace}
\newcommand{\fivethreesixnn}{$\sqrt{s_{\mathrm{NN}}}~=~5.36$~Te\kern-.1emV\xspace}
\newcommand{\ninesixtwonn}{$\sqrt{s_{\mathrm{NN}}}~=~9.62$~Te\kern-.1emV\xspace}
\newcommand{\twoHnn}       {$\sqrt{s_{\text{\tiny NN}}}=200$~Ge\kern-.1emV\xspace}
\newcommand{\LT}           {L{\'e}vy-Tsallis\xspace}
\DeclareSIUnit{\csym}{\text{\ensuremath{c}}}
\DeclareSIUnit{\GeVc}{\giga\electronvolt\per\csym}
\newcommand{\MeVc}         {Me\kern-.1emV/$c$\xspace}
\newcommand{\GeVmass}      {Ge\kern-.2emV/$c^2$\xspace}
\newcommand{\MeVmass}      {Me\kern-.2emV/$c^2$\xspace}
\newcommand{\lumi}         {\ensuremath{\mathcal{L}_{\rm int}}\xspace}
\newcommand{\Mgg}          {\ensuremath{M_{\gamma\gamma}}\xspace}
\newcommand{\ShowerShape}  {\ensuremath{\sigma^2_\text{long}}\xspace}
\newcommand{\MC}  {MC\xspace}

\newcommand{\ITS}          {\rm{\ac{ITS}}\xspace}
\newcommand{\TOF}          {\rm{\ac{TOF}}\xspace}
\newcommand{\ZDC}          {\rm{\ac{ZDC}}\xspace}
\newcommand{\ZDCs}         {\rm{ZDCs}\xspace}
\newcommand{\ZNA}          {\rm{ZNA}\xspace}
\newcommand{\ZNC}          {\rm{ZNC}\xspace}
\newcommand{\SPD}          {\rm{\ac{SPD}}\xspace}
\newcommand{\SDD}          {\rm{\ac{SDD}}\xspace}
\newcommand{\SSD}          {\rm{\ac{SSD}}\xspace}
\newcommand{\TPC}          {\rm{\ac{TPC}}\xspace}
\newcommand{\TRD}          {\rm{\ac{TRD}}\xspace}
\newcommand{\VZERO}        {\rm{V$0$}\xspace}
\newcommand{\TZERO}        {\rm{T$0$}\xspace}
\newcommand{\TZEROA}        {\rm{T$0$A}\xspace}
\newcommand{\TZEROC}        {\rm{T$0$C}\xspace}
\newcommand{\VZEROA}       {\rm{V$0$A}\xspace}
\newcommand{\VZEROC}       {\rm{V$0$C}\xspace}
\newcommand{\Vdecay} 	   {\ensuremath{V^{0}}\xspace}

\newcommand{\ee}           {\ensuremath{e^{+}e^{-}}} 
\newcommand{\pip}          {\ensuremath{\pi^{+}}\xspace}
\newcommand{\pim}          {\ensuremath{\pi^{-}}\xspace}
\newcommand{\piz}          {\ensuremath{\pi^{0}}\xspace}
\newcommand{\pipipi}       {\ensuremath{\pi^{+}\pi^{-}\pi^{0}}\xspace}
\newcommand{\kap}          {\ensuremath{\rm{K}^{+}}\xspace}
\newcommand{\kam}          {\ensuremath{\rm{K}^{-}}\xspace}
\newcommand{\pbar}         {\ensuremath{\rm\overline{p}}\xspace}
\newcommand{\kzero}        {\ensuremath{{\rm K}^{0}_{\rm{S}}}\xspace}
\newcommand{\lmb}          {\ensuremath{\Lambda}\xspace}
\newcommand{\almb}         {\ensuremath{\overline{\Lambda}}\xspace}
\newcommand{\Om}           {\ensuremath{\Omega^-}\xspace}
\newcommand{\Mo}           {\ensuremath{\overline{\Omega}^+}\xspace}
\newcommand{\X}            {\ensuremath{\Xi^-}\xspace}
\newcommand{\Ix}           {\ensuremath{\overline{\Xi}^+}\xspace}
\newcommand{\Xis}          {\ensuremath{\Xi^{\pm}}\xspace}
\newcommand{\Oms}          {\ensuremath{\Omega^{\pm}}\xspace}
\newcommand{\etpr}          {\ensuremath{\eta/\pi^{0}}\rm{ ratio}\xspace}
\newcommand{\etp}          {\ensuremath{\eta/\pi^{0}}\xspace}

\newcommand{\SigUnityROO}{\ensuremath{9.2\,\sigma}\xspace}
\newcommand{\SigEPPSROO}{\ensuremath{1.8\,\sigma}\xspace}
\newcommand{\SigNNNPDFDoubleRatio}{\ensuremath{4.9\,\sigma}\xspace}

\newcommand{\reffig}[1]{\hyperref[fig:#1]{Fig.~\ref*{fig:#1}}}
\newcommand{\refsec}[1]{\hyperref[s:#1]{Sec.~\ref*{s:#1}}}
\newcommand{\refeq}[1]{\hyperref[eq:#1]{Eq.~\ref*{eq:#1}}}
\newcommand{\reftab}[1]{\hyperref[tab:#1]{Tab.~\ref*{tab:#1}}}
\newcommand{\refFig}[1]{\hyperref[fig:#1]{Figure~\ref*{fig:#1}}}
\newcommand{\refSec}[1]{\hyperref[s:#1]{Section~\ref*{s:#1}}}
\newcommand{\refTab}[1]{\hyperref[tab:#1]{Table~\ref*{tab:#1}}}
\newcommand{\comment}[1]{\color{blue}#1\color{black}}

\newcommand{\ToDo}[1]{\textcolor{red}{#1}}
\begin{titlepage}
\PHyear{2026}       
\PHnumber{171}      
\PHdate{8 June}  

\title{Evidence for parton energy loss\\in oxygen--oxygen collisions at 
$\sqrt{s_{\text{\footnotesize NN}}}$~=~5.36~Te\kern-.1emV}

\ShortTitle{Evidence for parton energy loss in oxygen-oxygen collisions}   

\Collaboration{ALICE Collaboration\thanks{See Appendix~\ref{app:collab} for the list of collaboration members}}
\ShortAuthor{ALICE Collaboration} 

\begin{abstract}
Ultra-relativistic heavy-ion collisions create a hot and dense medium of deconfined quarks and gluons, the quark--gluon plasma~(QGP), in which parton energy loss (``jet quenching'') is a key probe of hot medium properties. 
While parton energy loss has been firmly established in large systems such as Pb--Pb and Au--Au collisions, no unambiguous direct evidence exists in smaller systems such as high-multiplicity p--Pb and pp collisions. 
To probe the onset of parton energy loss at intermediate system size, measurements of neutral-pion production are presented in this Letter for oxygen--oxygen (OO) and proton--oxygen (pO) collisions recorded with the ALICE detector in July 2025, relative to a pp baseline.
The nuclear modification factor \ROO is suppressed relative to unity with a transverse-momentum dependence similar to that observed in Pb--Pb collisions, consistent with a previous CMS measurement in OO collisions with charged particles.
As \ROO contains contributions from both cold and hot nuclear matter effects, \RpO is also presented in order to constrain cold nuclear matter~(CNM) contributions.
\RpO is found to be compatible with unity, indicating that CNM effects alone cannot account for the suppression observed in \ROO.
Final-state effects are isolated using the measured double ratio \ROOpOpO, which largely cancels CNM contributions and exhibits a significant suppression relative to expectations without energy loss at a \SigNNNPDFDoubleRatio level. 
Theoretical models incorporating parton energy loss via different mechanisms predict a significant suppression of the \ROOpOpO relative to unity, consistent with the data.
These findings establish parton energy loss in OO collisions, extending experimental evidence for jet quenching to the smallest nuclear system studied to date.

\end{abstract}
\end{titlepage}

\setcounter{page}{2} 



At very high temperatures and densities, strongly-interacting matter undergoes a phase transition to a state of matter consisting of deconfined quarks and gluons: the quark--gluon plasma (QGP).
A wide range of measurements in heavy-ion collisions at the Super Proton Synchrotron (SPS), Relativistic Heavy-Ion Collider (RHIC) and Large Hadron Collider (LHC) have provided compelling evidence for the formation of the QGP in Pb--Pb and Au--Au collisions, where the QGP behaves like an almost perfect liquid of strongly coupled quarks and gluons~\cite{WA98:2001lnw,dEnterria:2004cly,Busza:2018rrf,BRAHMS:2004adc,PHENIX:2004vcz,PHOBOS:2004zne,STAR:2005gfr,ALICE:2022wpn,CMS:2024krd}.
Hard scattering processes of quarks and gluons in hadronic collisions produce high-momentum partons with large virtuality that decay in a parton shower to observable hadrons.
Due to the short wavelength of the scattered partons, they interact microscopically with the quarks and gluons of the QGP and lose energy that gets transferred to the medium ("jet quenching")~\cite{Bjorken:1982tu,Wang:2025lct}.
The scattered partons, as well as the produced hadrons, are therefore sensitive to the microscopic structure of the QGP and its transport properties.
In order to study these effects, hadron and jet production are measured with respect to a pp reference via nuclear modification factors defined for nucleus-nucleus and proton-nucleus collisions as
\begin{equation}
\label{eq:RAA}
R_{\mathrm{AA}} =
\frac{1}{A^{2}}
\frac{\left.\mathrm{d}\sigma_{\mathrm{AA}}/\mathrm{d}p_{\mathrm{T}}\right.}
     {\left.\mathrm{d}\sigma_{\mathrm{pp}}/\mathrm{d}p_{\mathrm{T}}\right.},
\quad
R_{\mathrm{pA}} =
\frac{1}{A}
\frac{\left.\mathrm{d}\sigma_{\mathrm{pA}}/\mathrm{d}p_{\mathrm{T}}\right.}
     {\left.\mathrm{d}\sigma_{\mathrm{pp}}/\mathrm{d}p_{\mathrm{T}}\right.},
\end{equation}
where \(\sigma\) denotes production cross sections in minimum-bias collisions and \(A\) is the nuclear mass number.
Strong suppression ($\RAA \ll 1$) of inclusive charged hadrons, identified light-flavor hadrons, as well as inclusive and semi-inclusive jets, has been observed at RHIC and the LHC~\cite{STAR:2020xiv,ATLAS:2010isq,ALICE:2010yje,CMS:2011iwn}, showcasing significant energy loss in central (head-on) heavy-ion collisions.
The formation of the QGP in Pb--Pb and Au--Au collisions is further corroborated by the observation of collective flow phenomena, such as the anisotropic flow coefficient $v_{2}$ and higher harmonics~\cite{VOLOSHIN2012541}.
Small systems, such as proton-proton~(pp) and proton-lead~(p--Pb), have traditionally been considered ``QGP-free'' references for Pb--Pb collisions.
These assumptions have been challenged in the past decade as several features indicating collectivity, such as long-range correlations and strangeness enhancement, were observed in high-multiplicity pp and p--Pb collisions~\cite{CMSRidgepp,ALICE:2012eyl,ALICE:2016fzo,ALICE:2024vzv}. 
However, one key signature of QGP formation remains elusive in these collision systems: jet quenching~\cite{Wang:2025lct}.
Establishing the presence or absence of parton energy loss in progressively smaller nuclear collision systems is therefore a key step toward understanding the onset of QGP formation at low mass numbers.

In July 2025, the LHC delivered its first oxygen--oxygen~(OO) and proton--oxygen~(pO) collisions, providing an intermediate system size that bridges the gap between small (pp and p--Pb) and large (Xe--Xe and Pb--Pb) collision systems. 
As observed in high-multiplicity pp and p--Pb collisions, initial results of anisotropic flow from OO collisions at the LHC ~\cite{ALICE:2025luc,ATLAS:2025nnt, CMS:2025tga} show clear signs of collectivity, where in particular the data is in good agreement with hydrodynamic calculations.
A recent CMS measurement reported significant suppression of the charged-particle \ROO~\cite{CMSROO}, suggesting possible energy loss in OO collisions. 
This has been further explored by STAR at RHIC energies via semi-inclusive jet measurements~\cite{STAR:2026nfy}.
However, such measurements can be affected by cold nuclear matter~(CNM) effects ~\cite{Gebhard:2024flv,Mazeliauskas:2025clt}, arising from modifications of the parton distribution functions in bound nuclei, such as gluon shadowing, or fully coherent energy loss of the scattered partons~\cite{Arleo:2020eia}.
Because CNM effects in oxygen are poorly constrained experimentally, their theoretical description via nuclear PDFs~(nPDFs) carries substantial uncertainties, complicating the interpretation of the observed suppression~\cite{Gebhard:2024flv,Mazeliauskas:2025clt}.
This Letter presents measurements of both \ROO and, for the first time, \RpO, providing unique constraints on CNM effects. 
To largely cancel CNM contributions and isolate hot-medium effects in OO collisions, the double ratio \ROOpOpO is introduced. 
This observable reduces the dominant theoretical uncertainties and enables a direct test of parton energy loss in oxygen--oxygen collisions~\cite{nPDFPredictionsPaper}.
These measurements are performed using neutral pions (\piz), which are abundantly produced at the LHC~\cite{ALICE:2024vgi} and, as leading jet fragments, are sensitive to parton energy loss in the QGP~\cite{ALICE:2018mdl}.
The results are obtained from OO collisions at \fivethreesixnn and from pO collisions at \ninesixtwonn, using pp reference data at \fivethreesix and an additional pp data set at \thirteensix to interpolate the baseline for the pO energy.

The analysis is based on data from the Fast Interaction Trigger (FIT)~\cite{ALICE:FIT} and the ALICE Electromagnetic Calorimeter (EMCal)~\cite{ALICE:EMCALPerf}. 
The FIT system provides the minimum-bias (MB)~trigger and initiates the EMCal readout used for photon reconstruction.
The relevant FIT components are the FT0-A and FT0-C, two segmented scintillator disks located at forward and backward rapidity. They provide an excellent resolution of the collision time of about $\SI{26}{ps}$.
Minimum-bias events are triggered by a coincident signal in both FT0-A and FT0-C.
The primary collision vertex is required to be within $\SI{10}{cm}$ from the nominal collision point in the beam direction.
The EMCal is a lead-scintillator sampling electromagnetic calorimeter, covering $\vert\eta\vert < 0.7$ and $\Delta \varphi = 173^{\circ}$. 
It comprises 17664 cells with granularity $\Delta\eta \approx \Delta\varphi \approx 0.0143$, enabling two-photon separation from \piz decays up to $\pt \approx \SI{20}{\GeVc}$.

The energy of a photon hitting the EMCal is typically distributed across multiple neighboring cells. 
In order to recover the full photon energy, adjacent cells are grouped into clusters, following the same approach as in Ref.~\cite{ALICE:2024vgi}. 
To enhance photon purity and reduce contributions from hadron-induced clusters, additional selections are applied, following the procedure outlined in Ref.~\cite{ALICE:EMCALPerf}. 
The clusters are required to have an energy $E_\text{clus} > \SI{600}{MeV}$. 
The photon purity in the cluster sample is further enhanced by a selection based on the shape of the shower using a parametrization of the shower surface ellipse axes.
In particular, round clusters are selected, as photons are expected to exhibit a round shape compared to hadronic clusters, which are more likely to be elongated.
The timing information of the EMCal cells in relation to the collision time is used to reject contributions from out-of-bunch pileup. 
With a bunch spacing of $\SI{25}{ns}$ and an EMCal timing resolution of $\sigma_\text{time} \approx\SI{2}{ns}$, clusters are required to satisfy $|t_\text{clus}| < \SI{15}{ns}$, largely eliminating out-of-bunch pileup.
The same selection criteria are applied to the clusters across all collision systems and collision energies, supported by stable detector conditions and the absence of significant multiplicity-dependent effects. 

The \piz mesons are reconstructed by calculating the invariant mass (\Mgg) and \pt of all cluster pairs within a single event. 
The resulting photon-pair distribution contains both the \piz signal and combinatorial background from pairs not originating from the same \piz. 
The background can be decomposed into uncorrelated and correlated components, the latter originating from pairs in the same decay chain. 
To extract the signal, the background is modeled as a linear combination of two components. The uncorrelated background is estimated using the mixed-event method~\cite{ALICE:2018vhm}, which combines clusters from different events to mimic uncorrelated pairs. The correlated background is estimated using the rotation method~\cite{ALICE:EMCALPerf}, which rotates one cluster in azimuth and therefore retains both correlated and uncorrelated contributions.
A template fit is used to determine the optimal weighted sum of the two background description methods.
After background subtraction, the \piz yield is obtained by integrating the counts around the \piz signal peak.

The production cross section in a transverse-momentum interval $\Delta$\pt and rapidity interval $\Delta y$ is defined as
\begin{equation}
    E \frac{\mbox{d}^3 \sigma}{\mbox{d}^3p} = \frac{1}{2\pi} \frac{1}{ \lumi } \frac{1}{A\varepsilon} \frac{1}{\mathcal{B}} \frac{N^{\piz} - N^{\piz}_{\rm sec}}{\pt \Delta y \Delta \pt},
    \label{eq:InvXsection}
\end{equation}
where $N^{\piz}$ is the raw yield and $N^{\piz}_{\rm sec}$ accounts for secondary \piz contributions to the yield from decays of long-lived particles ($c\tau > \SI{1}{cm}$) and interactions with detector material. 
These contributions amount to about $4\%$ of the yield, and are estimated using PYTHIA~8~\cite{PYTHIA_PhysicsAndUsage} and GEANT4~\cite{GEANT4:2002zbu}.
The EMCal acceptance ($A$) for \piz produced within $|y| < 0.8$ is about 0.4 at high \pt, and decreases at lower \pt due to the increasing opening angle of the $\piz\rightarrow\gamma\gamma$~decay. 
The reconstruction efficiency $\varepsilon$ is determined using \MC simulations with a GEANT4-based detector response and increases from about $5\%$ at low \pt to roughly $40\%$ at intermediate \pt, before decreasing again for $\pt \gtrsim \SI{12}{\GeVc}$ due to the increasing impact of cluster merging.
The spectra are normalized by the branching ratio ($\mathcal{B} = (98.823 \pm 0.034)\%$)~\cite{PDG}.
The analyzed luminosity is determined as $\lumi=N_{\rm evt}/\sigma_{\rm vis}^{\rm MB}$, where $N_{\rm evt}$ denotes the number of events, i.e., MB-triggered collisions recorded with the EMCal and $\sigma_{\rm vis}^{\rm MB}$ is the visible cross section of the MB trigger, measured in van der Meer scans for each data set, with the procedure outlined in Refs.~\cite{ALICE:2021leo,ALICE:2022xir}.

The number of analyzed collisions is obtained from the number of triggered bunch crossings by correcting for the average number of collisions per triggered bunch crossing based on the interaction rates. 
This pileup correction ranges from 1.2\% in pp to 4.5\% in OO collisions.
The resulting luminosities are {$\lumi^{\rm OO} = \SI{1.87\pm0.04}{nb^{-1}}$}, $\lumi^{\rm pO} = \SI{3.6\pm0.1}{nb^{-1}}$, {$\lumi^{\rm pp,\,5.36\,TeV} = \SI{22.8\pm0.6}{nb^{-1}}$} and {$\lumi^{\rm pp,\,13.6\,TeV} = \SI{137\pm4}{nb^{-1}}$}, with systematic uncertainties stemming from the determination of $\sigma_{\rm vis}^{\rm MB}$.
Effects of the finite \pt interval width are accounted for by shifting the points horizontally to the \pt where the values of the fitted, unbinned spectrum are equal to their means over the corresponding \pt intervals~\cite{BinShift}. 
For \RpO, \ROO, and \ROOpOpO, the corresponding shift is performed in the vertical direction according to the underlying spectra. This shift is negligible for \RpO and below $1\%$ for \ROO and \ROOpOpO.

The determination of \RpO using~\refeq{RAA} requires a measurement of the \piz production cross section at the same collision energy and rapidity interval as the \pO measurement. 
This reference is obtained by interpolating between pp measurements at \fivethreesix and \thirteensix.
The interpolation is performed by parameterizing the $\sqrt{s}$ dependence of the \piz cross section at LHC energies~\cite{ALICE:2014hpa,ALICE:2012wos,ALICE:2017ryd,ALICE:2024vgi} with a power-law function in each \pt interval.
The resulting parameterizations are used to scale the \fivethreesix data to \ninesixtwo and the \thirteensix data down to \ninesixtwo, yielding two independent cross sections, which are then averaged to obtain the final interpolated reference.
Additional approaches to describe the center-of-mass energy dependence are considered as part of the systematic uncertainties outlined below.
The two-in-one design of the LHC magnets requires equal magnetic rigidity for the proton and oxygen beams. This leads to different energies per nucleon and a boost of the nucleon--nucleon center-of-mass frame relative to the laboratory frame.
As a consequence, the measured rapidity in pO is $-1.15\,<\,y\,<\,0.45$, requiring the pp baseline to be shifted to the same rapidity interval. 
The rapidity shift correction, estimated using PYTHIA~8~\cite{PYTHIA_PhysicsAndUsage}, leads to a \pt-dependent reduction of the interpolated pp baseline by less than $\SI{1}{\%}$.

The systematic uncertainties due to the choice of cluster selections and the signal extraction parameters are evaluated by varying these selections and parameters and studying their impact on the cross sections and nuclear modification factors. 
For the cross sections, the leading sources of systematic uncertainty in the intermediate \pt range arise from the selections based on the cluster shape ($\sim1.5\%$), the energy calibration ($\sim2\%$), and the uncertainty of the material budget~($4.2\%$)~\cite{ALICE:2017ryd}.
The variations are performed simultaneously for all cross sections contributing to the nuclear modification factors \ROO, \RpO and \ROOpOpO, allowing correlated contributions to cancel. 
The uncertainty of the material budget, as well as most uncertainties related to the cluster selection, cancel in the nuclear modification factor. 
Consequently, the leading uncertainty arises from the energy calibration ($\sim3\%$). 
For \RpO and \ROOpOpO, the energy interpolation presents a dominant contribution with $\sim2\%$ and~$\sim4\%$, respectively. 
This uncertainty is evaluated by performing the energy interpolation with two alternative estimators for the energy dependence~(a deep neural network trained on charged hadron spectra~\cite{DNNInterpolationPaper} and pQCD calculations~\cite{nPDFPredictionsPaper}).
The different contributions add up to relative systematic uncertainties of $\sim5\%$ for the measured cross sections, $\sim3\%$ for \ROO and \RpO, and $\sim5\%$ for \ROOpOpO.
In addition, the measurement of the visible cross sections in the various systems via the van der Meer scans introduces global normalization uncertainties of the measured production cross sections of 2.6\% in pp at \fivethreesix, 2.6\% in pp at \thirteensix, 2.7\% in pO and 2\% in OO collisions. 
These normalization uncertainties are treated as uncorrelated in the nuclear modification factors and represented around unity in the following figures.


\refFig{XSecs} shows the measured \piz cross sections in pp collisions at \fivethirteen, as well as in pO collisions at \ninesixtwonn, and in OO collisions at \fivethreesixnn. 
All spectra cover a \pt interval of $1.2 < \pt < \SI{20}{\GeVc}$.
Additionally, the interpolated pp cross section at \ninesixtwo is shown, which is used as a reference for \RpO. 
Both the pO cross section as well as the pp cross section at \ninesixtwo are given in the center-of-mass rapidity interval of $-1.15\,<\,y\,<\,0.45$, whereas the other systems are reported in $|y|\,<\,0.8$.

\begin{figure}[th!]
         \centering
         \includegraphics[width=0.55\textwidth]{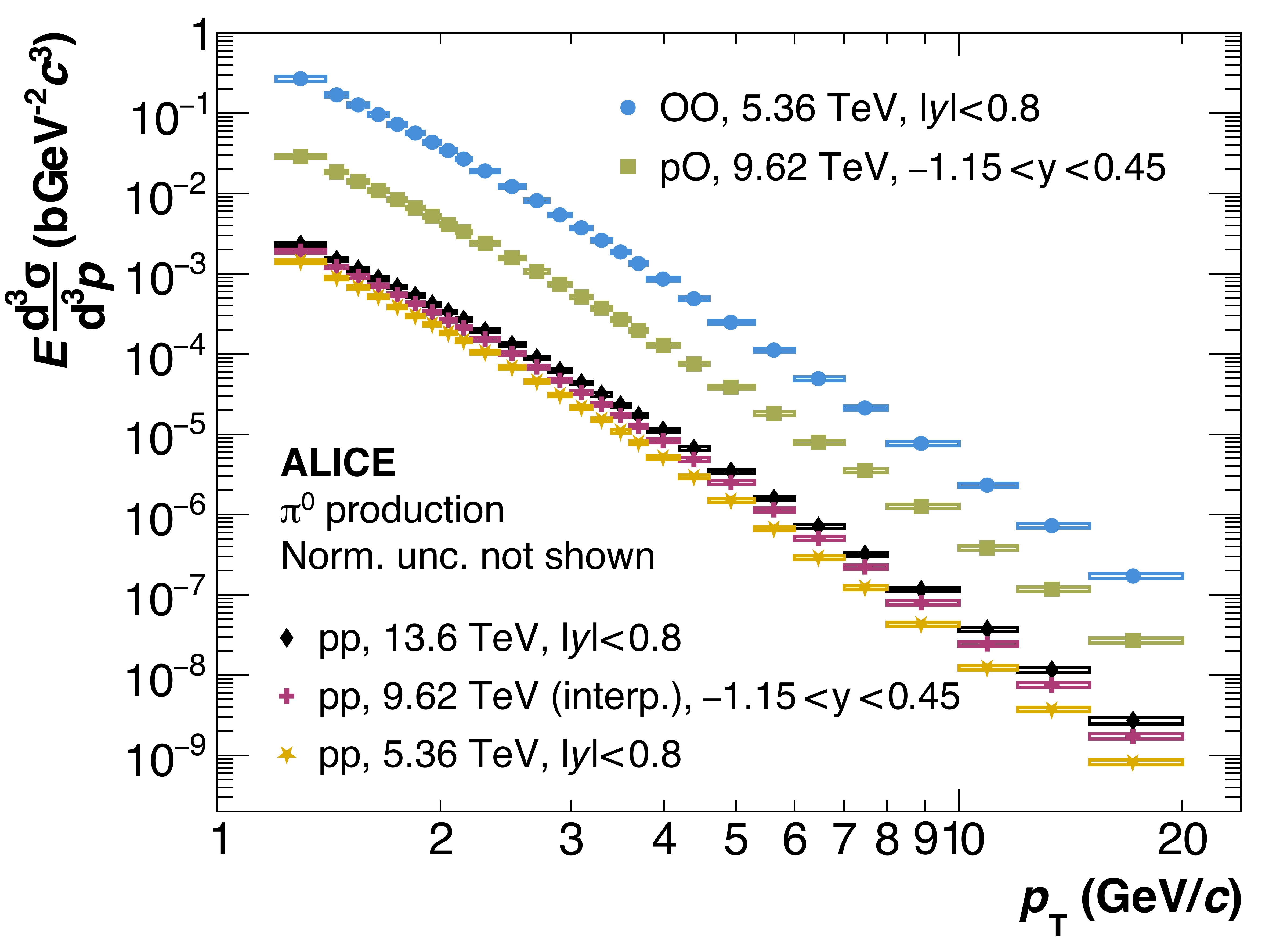}
        \caption{Production cross section of \piz measured in OO collisions at \fivethreesixnn, in pO collisions at \ninesixtwonn and in pp collisions at \fivethirteen, together with the interpolated pp reference at \ninesixtwo used for the \RpO calculation. Vertical bars (boxes) represent the statistical (systematic) uncertainty. Normalization uncertainties (not shown) originating from the luminosity determination are given in the text.}
        \label{fig:XSecs}
\end{figure}

\begin{figure}[th!]
         \centering
         \includegraphics[width=0.55\textwidth]{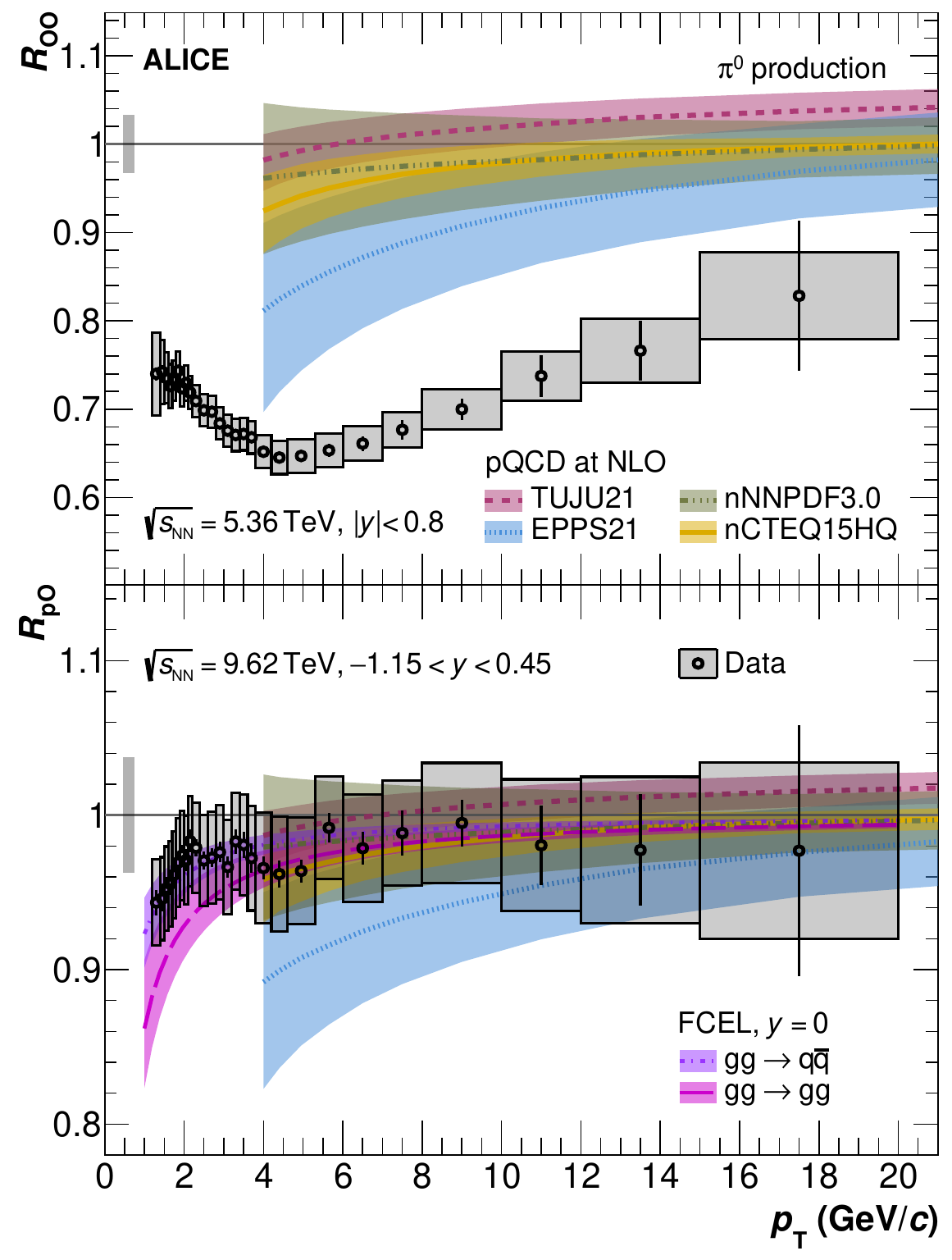}
        \caption{Nuclear modification factors \ROO (upper panel) and \RpO (lower panel). 
        Vertical bars (boxes) represent the statistical (systematic) uncertainty, while a box around unity denotes the normalization uncertainty due to the luminosity determination. Shaded bands show predictions~\cite{nPDFPredictionsPaper} of four nPDFs and their 68\% confidence level, including nPDF and scale uncertainty. The lower panel additionally includes two calculations of fully coherent energy loss (FCEL) for the two partonic channels gg$\rightarrow$q$\bar{\text{q}}$ and gg$\rightarrow$gg~\cite{Arleo:2020eia,Arleo:2020hat,Arleo:2021krm}.}
        \label{fig:NuclearModFactors}
\end{figure}

\refFig{NuclearModFactors} presents the nuclear modification factors \ROO and \RpO as a function of \pt, in the upper and lower panels, respectively. 
\ROO shows a clear suppression with respect to unity, with a significance of \SigUnityROO, in line with recent CMS results~\cite{CMSROO}.
The significance is evaluated over the \pt range covered by the corresponding prediction using pseudo-experiments generated according to the null hypothesis and the full covariance matrix of data and model uncertainties. For each pseudo-experiment, a pull quantifying the covariance-weighted suppression relative to the expectation is calculated and used as a test statistic, following the definition introduced in Ref.~\cite{Eskola:2022rlm}. The $p$-value, from which the significance is derived, is determined from the fraction of pseudo-experiments yielding a larger pull than that observed in data. The same procedure is used for all significances quoted in this Letter.
The data are compared with four different pQCD calculations at NLO using recent nPDFs that incorporate CNM effects~\cite{incnlowebsite,INCNLO1,INCNLO2,nPDFPredictionsPaper}.
While the calculations using TUJU21~\cite{Helenius:2021tof}, nNNPDF3.0~\cite{AbdulKhalek:2022fyi}, and nCTEQ15HQ~\cite{Duwentaster:2022kpv} show deviations of at most about $5\,\%$ from unity, the prediction using EPPS21~\cite{Eskola:2021nhw} shows a deviation from unity of up to $20\,\%$. 
Given the sizable nPDF uncertainties, shown as shaded bands, a suppression of \piz production in OO collisions beyond theoretical CNM expectations can only be established with a significance of \SigEPPSROO when compared to the most suppressed prediction using EPPS21.

The lower panel shows the measured \RpO, which provides, for the first time, a data-driven reference for CNM effects in an Oxygen projectile. 
The data are compatible with unity over the full \pt range, indicating that CNM effects lead to no significant suppression.
Within the uncertainties, the data are compatible with pQCD calculations for all nPDFs.
In addition, calculations based solely on fully coherent energy loss (FCEL)~\cite{Arleo:2020eia,Arleo:2020hat,Arleo:2021krm} are shown.
The calculations describe the data within uncertainties, highlighting that an alternative theoretical formalism to describe CNM effects is also consistent with the measurement.

\begin{figure}
         \centering
         \includegraphics[width=0.55\textwidth]{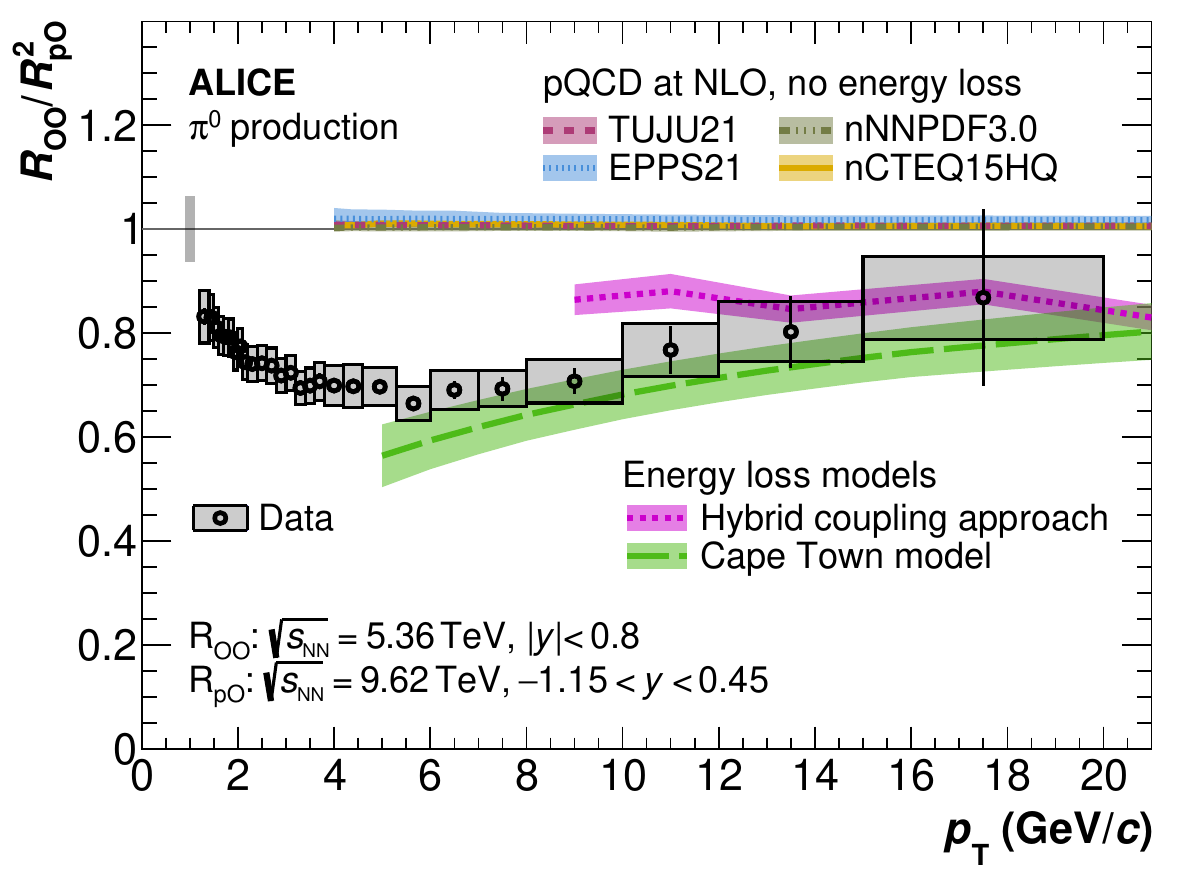}
         \caption{Double ratio \ROOpOpO compared to pQCD calculations including only CNM effects~\cite{nPDFPredictionsPaper}, located around unity, and to energy loss models~\cite{ELossHybrid, ELossCollRad}, which exhibit a suppression similar to the one observed in data. Vertical bars (boxes) represent the statistical (systematic) uncertainty, while a box around unity denotes the normalization uncertainty due to the luminosity determination.}
        \label{fig:ELossTheory}
\end{figure}

To fully exploit these experimental findings, the double ratio \ROOpOpO is constructed to largely cancel CNM effects present in \ROO, as suggested in Ref.~\cite{nPDFPredictionsPaper}. 
The measured \ROOpOpO is shown in~\reffig{ELossTheory}, together with the corresponding pQCD calculations~\cite{nPDFPredictionsPaper} incorporating CNM effects.
These predictions are close to unity, reflecting the cancellation of CNM contributions and their associated nPDF uncertainties in the double ratio despite the differing collision energies and rapidity intervals of \ROO and \RpO.
The measured \ROOpOpO deviates significantly from all of these expectations without energy loss, with a significance of \SigNNNPDFDoubleRatio for nNNPDF3.0, where slightly larger significances are obtained for the other nPDF predictions.
This demonstrates that the observed modification cannot be explained by CNM effects alone and establishes the presence of parton energy loss in OO collisions.

Building on these findings, the data in~\reffig{ELossTheory} are confronted with two theoretical models that incorporate parton energy loss in a hot medium.
In the hybrid coupling approach~\cite{ELossHybrid,Kudinoor:2026wcs} the hard jet evolution is treated perturbatively. 
Soft jet-medium interactions are modeled using an energy-loss prescription whose functional form is motivated by holographic calculations for strongly coupled plasmas, with an overall normalization constrained by experimental data. 
The calculation uses event-by-event hydrodynamic profiles to account for event-by-event fluctuations in the initial geometry of the formed QGP droplets~\cite{Mantysaari:2025tcg}.
The Cape Town model~\cite{ELossCollRad} includes both collisional and radiative energy loss in the medium using a pQCD-based prescription, and free parameters are constrained by heavy-ion data from LHC and RHIC.
Both models predict a significant suppression of this observable due to parton energy loss, describing both the magnitude and shape of the suppression observed in data within the uncertainties.
This corroborates the experimental evidence of parton energy loss in OO collisions, showing that models incorporating energy loss more accurately describe the observed suppression, while those including only CNM effects fail to do so.

In summary, \piz production cross sections are measured in pp collisions at \fivethirteen, in pO collisions at \ninesixtwonn, and in OO collisions at \fivethreesixnn within the interval $1.2 < \pt < \SI{20}{\GeVc}$.
The nuclear modification factor \ROO is significantly suppressed relative to unity, with a transverse-momentum dependence similar to that observed in Pb--Pb collisions.
To constrain cold nuclear matter effects, \RpO is presented for the first time and is found to be consistent with unity within uncertainties.
Final-state effects are isolated using the double ratio \ROOpOpO, which largely cancels CNM contributions and demonstrates a significant suppression relative to expectations without energy loss at the level of \SigNNNPDFDoubleRatio.
The suppression is consistent with theoretical models that assume the presence of a hot medium.
These findings provide strong evidence for parton energy loss in the smallest nuclear collision system studied to date.

Supplemental material for this Letter is provided in Ref.~\cite{ALICE-PUBLIC-2026-002}.


\newenvironment{acknowledgement}{\relax}{\relax}
\begin{acknowledgement}
\section*{Acknowledgements}
We thank A. Mazeliauskas and P. Paakkinen for the useful discussions and input regarding the nPDF baseline predictions and their uncertainty determination. We also thank A. Kudinoor and C. Faraday for providing energy loss calculations, as well as G. Jackson for providing the FCEL predictions.

The ALICE Collaboration would like to thank all its engineers and technicians for their invaluable contributions to the construction of the experiment and the CERN accelerator teams for the outstanding performance of the LHC complex.
The ALICE Collaboration gratefully acknowledges the resources and support provided by all Grid centres and the Worldwide LHC Computing Grid (WLCG) collaboration.
The ALICE Collaboration acknowledges the following funding agencies for their support in building and running the ALICE detector:
A. I. Alikhanyan National Science Laboratory (Yerevan Physics Institute) Foundation (ANSL), State Committee of Science and World Federation of Scientists (WFS), Armenia;
Austrian Academy of Sciences, Austrian Science Fund (FWF): [M 2467-N36] and Nationalstiftung f\"{u}r Forschung, Technologie und Entwicklung, Austria;
Ministry of Communications and High Technologies, National Nuclear Research Center, Azerbaijan;
Rede Nacional de Física de Altas Energias (Renafae), Financiadora de Estudos e Projetos (Finep), Funda\c{c}\~{a}o de Amparo \`{a} Pesquisa do Estado de S\~{a}o Paulo (FAPESP) and The Sao Paulo Research Foundation  (FAPESP), Brazil;
Bulgarian Ministry of Education and Science, within the National Roadmap for Research Infrastructures 2020-2027 (object CERN), Bulgaria;
Ministry of Education of China (MOEC) , Ministry of Science \& Technology of China (MSTC) and National Natural Science Foundation of China (NSFC), China;
Ministry of Science and Education and Croatian Science Foundation, Croatia;
Centro de Aplicaciones Tecnol\'{o}gicas y Desarrollo Nuclear (CEADEN), Cubaenerg\'{\i}a, Cuba;
Ministry of Education, Youth and Sports of the Czech Republic, Czech Republic;
The Danish Council for Independent Research | Natural Sciences, the VILLUM FONDEN and Danish National Research Foundation (DNRF), Denmark;
Helsinki Institute of Physics (HIP), Finland;
Commissariat \`{a} l'Energie Atomique (CEA) and Institut National de Physique Nucl\'{e}aire et de Physique des Particules (IN2P3) and Centre National de la Recherche Scientifique (CNRS), France;
Bundesministerium f\"{u}r Forschung, Technologie und Raumfahrt (BMFTR) and GSI Helmholtzzentrum f\"{u}r Schwerionenforschung GmbH, Germany;
National Research, Development and Innovation Office, Hungary;
Department of Atomic Energy Government of India (DAE), Department of Science and Technology, Government of India (DST), University Grants Commission, Government of India (UGC) and Council of Scientific and Industrial Research (CSIR), India;
National Research and Innovation Agency - BRIN, Indonesia;
Istituto Nazionale di Fisica Nucleare (INFN), Italy;
Japanese Ministry of Education, Culture, Sports, Science and Technology (MEXT) and Japan Society for the Promotion of Science (JSPS) KAKENHI, Japan;
Consejo Nacional de Ciencia (CONACYT) y Tecnolog\'{i}a, through Fondo de Cooperaci\'{o}n Internacional en Ciencia y Tecnolog\'{i}a (FONCICYT) and Direcci\'{o}n General de Asuntos del Personal Academico (DGAPA), Mexico;
Nederlandse Organisatie voor Wetenschappelijk Onderzoek (NWO), Netherlands;
The Research Council of Norway, Norway;
Pontificia Universidad Cat\'{o}lica del Per\'{u}, Peru;
Ministry of Science and Higher Education, National Science Centre and WUT ID-UB, Poland;
Korea Institute of Science and Technology Information and National Research Foundation of Korea (NRF), Republic of Korea;
Ministry of Education and Scientific Research, Institute of Atomic Physics, Ministry of Research and Innovation and Institute of Atomic Physics and Universitatea Nationala de Stiinta si Tehnologie Politehnica Bucuresti, Romania;
Ministerstvo skolstva, vyskumu, vyvoja a mladeze SR, Slovakia;
National Research Foundation of South Africa, South Africa;
Swedish Research Council (VR) and Knut \& Alice Wallenberg Foundation (KAW), Sweden;
European Organization for Nuclear Research, Switzerland;
Suranaree University of Technology (SUT), National Science and Technology Development Agency (NSTDA) and National Science, Research and Innovation Fund (NSRF via PMU-B B05F650021), Thailand;
Turkish Energy, Nuclear and Mineral Research Agency (TENMAK), Turkey;
National Academy of  Sciences of Ukraine, Ukraine;
Science and Technology Facilities Council (STFC), United Kingdom;
National Science Foundation of the United States of America (NSF) and United States Department of Energy, Office of Nuclear Physics (DOE NP), United States of America.
In addition, individual groups or members have received support from:
Czech Science Foundation (grant no. 23-07499S), Czech Republic;
FORTE project, reg.\ no.\ CZ.02.01.01/00/22\_008/0004632, Czech Republic, co-funded by the European Union, Czech Republic;
European Research Council (grant nos. 101220549, 950692), European Union;
Deutsche Forschungs Gemeinschaft (DFG, German Research Foundation) ``Neutrinos and Dark Matter in Astro- and Particle Physics'' (grant no. SFB 1258), Germany;
CONVECS project, CUP C97H23001700002 FESR 2021-2027 program, Italy.

\end{acknowledgement}

\bibliographystyle{utphys}   
\bibliography{bibliography}

@article{ALICE:FIT,
    author = "Slupecki, Maciej",
    collaboration = "ALICE",
    title = "{Fast Interaction Trigger for ALICE upgrade}",
    doi = "10.1016/j.nima.2022.167021",
    journal = "Nucl. Instrum. Meth. A",
    volume = "1039",
    pages = "167021",
    year = "2022"
}

@article{STAR:2026nfy,
    collaboration = "STAR",
    title = "{Measurement of jet quenching in O+O collisions at $\sqrt{s_\mathrm{NN}}=200$ GeV by the STAR experiment at RHIC}",
    eprint = "2604.13935",
    archivePrefix = "arXiv",
    primaryClass = "nucl-ex",
    month = "4",
    year = "2026"
}

@article{ALICE:2025luc,
    author = "Abualrob, Ibrahim Jaser and others",
    collaboration = "ALICE",
    title = "{Evidence of nuclear geometry-driven anisotropic flow in OO and Ne$-$Ne collisions at $\mathbf{\sqrt{{\textit s}_{\rm\mathbf {NN}}}}$ = 5.36 TeV}",
    eprint = "2509.06428",
    archivePrefix = "arXiv",
    primaryClass = "nucl-ex",
    reportNumber = "CERN-EP-2025-203",
    month = "9",
    year = "2025"
}

@article{ATLAS:2025nnt,
    author = "Aad, Georges and others",
    collaboration = "ATLAS",
    title = "{Measurement of the azimuthal anisotropy of charged particles in $\mathbf{\sqrt{{\textit s}_{\rm\mathbf {NN}}}} = 5.36~{}^{16}\text{O}+{}^{16}\text{O}$ and ${}^{20}\text{Ne}+{}^{20}\text{Ne}$ collisions with the ATLAS detector}",
    eprint = "2509.05171",
    archivePrefix = "arXiv",
    primaryClass = "nucl-ex",
    reportNumber = "CERN-EP-2025-200",
    doi = "10.1103/xqxz-8bhf",
    journal = "Phys. Rev. C",
    volume = "113",
    number = "4",
    pages = "045205",
    year = "2026"
}

@article{CMS:2025tga,
    author = "Hayrapetyan, Aram and others",
    collaboration = "CMS",
    title = "{Observation of long-range collective flow in OO and NeNe collisions and implications for nuclear structure studies}",
    eprint = "2510.02580",
    archivePrefix = "arXiv",
    primaryClass = "nucl-ex",
    reportNumber = "CMS-HIN-25-009, CERN-EP-2025-222",
    month = "10",
    year = "2025"
}

@article{GEANT4:2002zbu,
    author = "Agostinelli, S. and others",
    collaboration = "GEANT4",
    title = "{GEANT4 - A Simulation Toolkit}",
    reportNumber = "SLAC-PUB-9350, FERMILAB-PUB-03-339, CERN-IT-2002-003",
    doi = "10.1016/S0168-9002(03)01368-8",
    journal = "Nucl. Instrum. Meth. A",
    volume = "506",
    pages = "250--303",
    year = "2003"
}

@article{ALICE:2024vgi,
    author = "Acharya, Shreyasi and others",
    collaboration = "ALICE",
    title = "{Light neutral-meson production in pp collisions at $\sqrt{s}$ = 13 TeV}",
    eprint = "2411.09560",
    archivePrefix = "arXiv",
    primaryClass = "hep-ex",
    reportNumber = "CERN-EP-2024-304",
    doi = "10.1007/JHEP08(2025)035",
    journal = "JHEP",
    volume = "08",
    pages = "035",
    year = "2025"
}

@article{ALICE:2022wpn,
    author = "Acharya, Shreyasi and others",
    collaboration = "ALICE",
    title = "{The ALICE experiment: a journey through QCD}",
    eprint = "2211.04384",
    archivePrefix = "arXiv",
    primaryClass = "nucl-ex",
    reportNumber = "CERN-EP-2022-227",
    doi = "10.1140/epjc/s10052-024-12935-y",
    journal = "Eur. Phys. J. C",
    volume = "84",
    number = "8",
    pages = "813",
    year = "2024"
}

@article{ALICE:EMCALPerf,
    author = "Acharya, Shreyasi and others",
    collaboration = "ALICE",
    title = "{Performance of the ALICE Electromagnetic Calorimeter}",
    eprint = "2209.04216",
    archivePrefix = "arXiv",
    primaryClass = "physics.ins-det",
    reportNumber = "CERN-EP-2022-184",
    doi = "10.1088/1748-0221/18/08/P08007",
    journal = "JINST",
    volume = "18",
    number = "08",
    pages = "P08007",
    year = "2023"
}

@article{ALICE:2022xir,
    author = "Acharya, Shreyasi and others",
    collaboration = "ALICE",
    title = "{ALICE luminosity determination for Pb$-$Pb collisions at $\sqrt{s_{\mathrm{NN}}} = 5.02$ TeV}",
    eprint = "2204.10148",
    archivePrefix = "arXiv",
    primaryClass = "nucl-ex",
    reportNumber = "CERN-EP-2022-072",
    doi = "10.1088/1748-0221/19/02/P02039",
    journal = "JINST",
    volume = "19",
    number = "02",
    pages = "P02039",
    year = "2024"
}

@article{ALICE:2018mdl,
    author = "Acharya, Shreyasi and others",
    collaboration = "ALICE",
    title = "{Neutral pion and $\eta$ meson production at mid-rapidity in Pb-Pb collisions at $\sqrt{s_{NN}}$ = 2.76 TeV}",
    eprint = "1803.05490",
    archivePrefix = "arXiv",
    primaryClass = "nucl-ex",
    reportNumber = "CERN-EP-2018-040",
    doi = "10.1103/PhysRevC.98.044901",
    journal = "Phys. Rev. C",
    volume = "98",
    number = "4",
    pages = "044901",
    year = "2018"
}

@article{ALICE:2018vhm,
    author = "Acharya, Shreyasi and others",
    collaboration = "ALICE",
    title = "{Neutral pion and $\eta$ meson production in p-Pb collisions at $\sqrt{s_\mathrm{NN}} = 5.02$ TeV}",
    eprint = "1801.07051",
    archivePrefix = "arXiv",
    primaryClass = "nucl-ex",
    reportNumber = "CERN-EP-2018-002",
    doi = "10.1140/epjc/s10052-018-6013-8",
    journal = "Eur. Phys. J. C",
    volume = "78",
    number = "8",
    pages = "624",
    year = "2018"
}

@article{ALICE:2017ryd,
    author = "Acharya, Shreyasi and others",
    collaboration = "ALICE",
    title = "{$\pi ^{0}$ and $\eta $ meson production in proton-proton collisions at $\sqrt{s}=8$ TeV}",
    eprint = "1708.08745",
    archivePrefix = "arXiv",
    primaryClass = "hep-ex",
    reportNumber = "CERN-EP-2017-216",
    doi = "10.1140/epjc/s10052-018-5612-8",
    journal = "Eur. Phys. J. C",
    volume = "78",
    number = "3",
    pages = "263",
    year = "2018"
}

@article{ALICE:2016fzo,
    author = "Adam, Jaroslav and others",
    collaboration = "ALICE",
    title = "{Enhanced production of multi-strange hadrons in high-multiplicity proton-proton collisions}",
    eprint = "1606.07424",
    archivePrefix = "arXiv",
    primaryClass = "nucl-ex",
    reportNumber = "CERN-EP-2016-153",
    doi = "10.1038/nphys4111",
    journal = "Nature Phys.",
    volume = "13",
    pages = "535--539",
    year = "2017"
}

@article{ALICE:2014hpa,
    author = "Abelev, Betty Bezverkhny and others",
    collaboration = "ALICE",
    title = "{Neutral pion production at midrapidity in pp and Pb-Pb collisions at $\sqrt{s_{{\mathrm {NN}}}}= 2.76\,{\mathrm {TeV}}$}",
    eprint = "1405.3794",
    archivePrefix = "arXiv",
    primaryClass = "nucl-ex",
    reportNumber = "CERN-PH-EP-2014-091",
    doi = "10.1140/epjc/s10052-014-3108-8",
    journal = "Eur. Phys. J. C",
    volume = "74",
    number = "10",
    pages = "3108",
    year = "2014"
}

@article{ALICE:2012eyl,
    author = "Abelev, Betty and others",
    collaboration = "ALICE",
    title = "{Long-range angular correlations on the near and away side in $p$-Pb collisions at $\sqrt{s_{NN}}=5.02$ TeV}",
    eprint = "1212.2001",
    archivePrefix = "arXiv",
    primaryClass = "nucl-ex",
    reportNumber = "CERN-PH-EP-2012-359",
    doi = "10.1016/j.physletb.2013.01.012",
    journal = "Phys. Lett. B",
    volume = "719",
    pages = "29--41",
    year = "2013"
}

@article{ALICE:2012wos,
    author = "Abelev, B. and others",
    collaboration = "ALICE",
    title = "{Neutral pion and $\eta$ meson production in proton-proton collisions at $\sqrt{s}=0.9$ TeV and $\sqrt{s}=7$ TeV}",
    eprint = "1205.5724",
    archivePrefix = "arXiv",
    primaryClass = "hep-ex",
    reportNumber = "CERN-PH-EP-2012-001",
    doi = "10.1016/j.physletb.2012.09.015",
    journal = "Phys. Lett. B",
    volume = "717",
    pages = "162--172",
    year = "2012"
}

@article{ALICE:2010yje,
    author = "Aamodt, K. and others",
    collaboration = "ALICE",
    title = "{Suppression of Charged Particle Production at Large Transverse Momentum in Central Pb-Pb Collisions at $\sqrt{s_{NN}} =$ 2.76 TeV}",
    eprint = "1012.1004",
    archivePrefix = "arXiv",
    primaryClass = "nucl-ex",
    reportNumber = "CERN-PH-EP-ALICE-2010-004, CERN-PH-EP-2010-066",
    doi = "10.1016/j.physletb.2010.12.020",
    journal = "Phys. Lett. B",
    volume = "696",
    pages = "30--39",
    year = "2011"
}

@article{CMSROO,
    author = "Hayrapetyan, Aram and others",
    collaboration = "CMS",
    title = "{Observation of Suppressed Charged-Particle Production in Ultrarelativistic Oxygen-Oxygen Collisions}",
    eprint = "2510.09864",
    archivePrefix = "arXiv",
    primaryClass = "nucl-ex",
    reportNumber = "CMS-HIN-25-008, CERN-EP-2025-226",
    doi = "10.1103/89sf-9t1x",
    journal = "Phys. Rev. Lett.",
    volume = "136",
    number = "16",
    pages = "162301",
    year = "2026"
}

@article{PDG,
    author = "Navas, S. and others",
    collaboration = "Particle Data Group",
    title = "{Review of particle physics}",
    doi = "10.1103/PhysRevD.110.030001",
    journal = "Phys. Rev. D",
    volume = "110",
    number = "3",
    pages = "030001",
    year = "2024"
}

@article{PYTHIA_PhysicsAndUsage,
    author = "Bierlich, Christian and others",
    title = "{A comprehensive guide to the physics and usage of PYTHIA 8.3}",
    eprint = "2203.11601",
    archivePrefix = "arXiv",
    primaryClass = "hep-ph",
    reportNumber = "LU-TP 22-16, MCNET-22-04, FERMILAB-PUB-22-227-SCD",
    doi = "10.21468/SciPostPhysCodeb.8",
    journal = "SciPost Phys. Codeb.",
    volume = "2022",
    pages = "8",
    year = "2022"
}

@article{BinShift,
    author = "Lafferty, G. D. and Wyatt, T. R.",
    title = "{Where to stick your data points: The treatment of measurements within wide bins}",
    reportNumber = "MAN-HEP-94-3, CERN-PPE-94-72, CERN-PPE-94-072",
    doi = "10.1016/0168-9002(94)01112-5",
    journal = "Nucl. Instrum. Meth. A",
    volume = "355",
    pages = "541--547",
    year = "1995"
}

@article{ELossHybrid,
    author = "Casalderrey-Solana, Jorge and Gulhan, Doga Can and Milhano, Jose Guilherme and Pablos, Daniel and Rajagopal, Krishna",
    title = "{Erratum to: A hybrid strong/weak coupling approach to jet quenching}",
    doi = "10.1007/JHEP09(2015)175",
    journal = "JHEP",
    volume = "2015",
    pages = "175",
    year = "2015"
}

@article{ELossCollRad,
    author = "Faraday, Coleridge and Horowitz, W. A.",
    title = "{Statistical analysis of pQCD energy loss across system size, flavor, $ \sqrt{s_{NN}} $, and p$_{T}$}",
    eprint = "2505.14568",
    archivePrefix = "arXiv",
    primaryClass = "hep-ph",
    doi = "10.1007/JHEP11(2025)019",
    journal = "JHEP",
    volume = "11",
    pages = "019",
    year = "2025"
}

@article{Busza:2018rrf,
    author = "Busza, Wit and Rajagopal, Krishna and van der Schee, Wilke",
    title = "{Heavy Ion Collisions: The big picture, and the big questions}",
    eprint = "1802.04801",
    archivePrefix = "arXiv",
    primaryClass = "hep-ph",
    reportNumber = "MIT-CTP-4892, MIT-CTP/4892",
    doi = "10.1146/annurev-nucl-101917-020852",
    journal = "Ann. Rev. Nucl. Part. Sci.",
    volume = "68",
    pages = "339--376",
    year = "2018"
}

@article{Helenius:2021tof,
    author = "Helenius, Ilkka and Walt, Marina and Vogelsang, Werner",
    title = "{NNLO nuclear parton distribution functions with electroweak-boson production data from the LHC}",
    eprint = "2112.11904",
    archivePrefix = "arXiv",
    primaryClass = "hep-ph",
    doi = "10.1103/PhysRevD.105.094031",
    journal = "Phys. Rev. D",
    volume = "105",
    number = "9",
    pages = "094031",
    year = "2022"
}

@article{Eskola:2021nhw,
    author = "Eskola, Kari J. and Paakkinen, Petja and Paukkunen, Hannu and Salgado, Carlos A.",
    title = "{EPPS21: a global QCD analysis of nuclear PDFs}",
    eprint = "2112.12462",
    archivePrefix = "arXiv",
    primaryClass = "hep-ph",
    doi = "10.1140/epjc/s10052-022-10359-0",
    journal = "Eur. Phys. J. C",
    volume = "82",
    number = "5",
    pages = "413",
    year = "2022"
}

@article{Duwentaster:2022kpv,
    author = {Duwent{\"a}ster, P. and Je{\v{z}}o, T. and Klasen, M. and Kova{\v{r}}{\'\i}k, K. and Kusina, A. and Muzakka, K. F. and Olness, F. I. and Ruiz, R. and Schienbein, I. and Yu, J. Y.},
    title = "{Impact of heavy quark and quarkonium data on nuclear gluon PDFs}",
    eprint = "2204.09982",
    archivePrefix = "arXiv",
    primaryClass = "hep-ph",
    reportNumber = "MS-TP-22-10, IFJPAN-IV-2022-5",
    doi = "10.1103/PhysRevD.105.114043",
    journal = "Phys. Rev. D",
    volume = "105",
    number = "11",
    pages = "114043",
    year = "2022"
}

@article{AbdulKhalek:2022fyi,
    author = "Abdul Khalek, Rabah and Gauld, Rhorry and Giani, Tommaso and Nocera, Emanuele R. and Rabemananjara, Tanjona R. and Rojo, Juan",
    title = "{nNNPDF3.0: evidence for a modified partonic structure in heavy nuclei}",
    eprint = "2201.12363",
    archivePrefix = "arXiv",
    primaryClass = "hep-ph",
    reportNumber = "Nikhef-2021-028, BONN-TH-2021-14",
    doi = "10.1140/epjc/s10052-022-10417-7",
    journal = "Eur. Phys. J. C",
    volume = "82",
    number = "6",
    pages = "507",
    year = "2022"
}

@article{Eskola:2022rlm,
    author = "Eskola, Kari J. and Paakkinen, Petja and Paukkunen, Hannu and Salgado, Carlos A.",
    title = "{Proton-PDF uncertainties in extracting nuclear PDFs from $W^\pm $ production in p+Pb collisions}",
    eprint = "2202.01074",
    archivePrefix = "arXiv",
    primaryClass = "hep-ph",
    doi = "10.1140/epjc/s10052-022-10179-2",
    journal = "Eur. Phys. J. C",
    volume = "82",
    number = "3",
    pages = "271",
    year = "2022"
}

@software{incnlowebsite,
  collaboration  = {INCNLO},
  author = "Aurenche, P. and Fontannaz, M. and Guillet, J. P. and Kniehl, Bernd A. and Werlen, M.",
  title   = {{INCNLO-direct photon and inclusive hadron production code - INCNLO version 1.4}},
  year    = {2002},
  urldate = {09/04/2023},
  version = {1.4},
  url     = {https://lapth.cnrs.fr/PHOX_FAMILY/readme_inc.html}
}

@article{INCNLO1,
    author = "Aversa, F. and Chiappetta, P. and Greco, Mario and Guillet, J. P.",
    title = "{QCD corrections to parton--parton scattering processes}",
    reportNumber = "CPT-88/P-2186, LNF-88/75-P",
    doi = "10.1016/0550-3213(89)90288-5",
    journal = "Nucl. Phys. B",
    volume = "327",
    pages = "105",
    year = "1989"
}

@article{INCNLO2,
    author = "Aurenche, P. and Fontannaz, M. and Guillet, J. P. and Kniehl, Bernd A. and Werlen, M.",
    title = "{Large $p_{\rm T}$ inclusive $\pi^0$ cross-sections and next-to-leading-order QCD predictions}",
    eprint = "hep-ph/9910252",
    archivePrefix = "arXiv",
    reportNumber = "LAPTH-751-99, LPT-ORSAY-99-78, DESY-99-153",
    doi = "10.1007/s100520000309",
    journal = "Eur. Phys. J. C",
    volume = "13",
    pages = "347--355",
    year = "2000"
}

@article{BRAHMS:2004adc,
    author = "Arsene, I. and others",
    collaboration = "BRAHMS",
    title = "{Quark gluon plasma and color glass condensate at RHIC? The Perspective from the BRAHMS experiment}",
    eprint = "nucl-ex/0410020",
    archivePrefix = "arXiv",
    doi = "10.1016/j.nuclphysa.2005.02.130",
    journal = "Nucl. Phys. A",
    volume = "757",
    pages = "1--27",
    year = "2005"
}

@article{PHOBOS:2004zne,
    author = "Back, B. B. and others",
    collaboration = "PHOBOS",
    title = "{The PHOBOS perspective on discoveries at RHIC}",
    eprint = "nucl-ex/0410022",
    archivePrefix = "arXiv",
    doi = "10.1016/j.nuclphysa.2005.03.084",
    journal = "Nucl. Phys. A",
    volume = "757",
    pages = "28--101",
    year = "2005"
}

@article{STAR:2005gfr,
    author = "Adams, John and others",
    collaboration = "STAR",
    title = "{Experimental and theoretical challenges in the search for the quark gluon plasma: The STAR Collaboration's critical assessment of the evidence from RHIC collisions}",
    eprint = "nucl-ex/0501009",
    archivePrefix = "arXiv",
    doi = "10.1016/j.nuclphysa.2005.03.085",
    journal = "Nucl. Phys. A",
    volume = "757",
    pages = "102--183",
    year = "2005"
}

@article{PHENIX:2004vcz,
    author = "Adcox, K. and others",
    collaboration = "PHENIX",
    title = "{Formation of dense partonic matter in relativistic nucleus--nucleus collisions at RHIC: Experimental evaluation by the PHENIX collaboration}",
    eprint = "nucl-ex/0410003",
    archivePrefix = "arXiv",
    doi = "10.1016/j.nuclphysa.2005.03.086",
    journal = "Nucl. Phys. A",
    volume = "757",
    pages = "184--283",
    year = "2005"
}

@article{CMS:2024krd,
    author = "Hayrapetyan, Aram and others",
    collaboration = "CMS",
    title = "{Overview of high-density QCD studies with the CMS experiment at the LHC}",
    eprint = "2405.10785",
    archivePrefix = "arXiv",
    primaryClass = "nucl-ex",
    reportNumber = "CMS-HIN-23-011, CERN-EP-2024-057",
    doi = "10.1016/j.physrep.2024.11.007",
    journal = "Phys. Rept.",
    volume = "1115",
    pages = "219--367",
    year = "2025"
}

@article{Wang:2025lct,
    author = "Wang, Xin-Nian and Wiedemann, Urs Achim",
    title = "{QGP@50: More than four decades of jet quenching}",
    eprint = "2508.18794",
    archivePrefix = "arXiv",
    primaryClass = "hep-ph",
    reportNumber = "CERN-TH-2025-173",
    month = "8",
    year = "2025"
}

@article{Bjorken:1982tu,
    author = "Bjorken, J. D.",
    title = "{Energy Loss of Energetic Partons in Quark - Gluon Plasma: Possible Extinction of High p(t) Jets in Hadron - Hadron Collisions}",
    journal = "FERMILAB-PUB-82-059-THY, FERMILAB-PUB-82-059-T",
    month = "8",
    year = "1982"
}

@article{VOLOSHIN2012541,
    author = "Voloshin, Sergei A.",
    editor = "Faessler, Amand and Rodin, Vadim",
    title = "{Collective phenomena in ultra-relativistic nuclear collisions: anisotropic flow and more}",
    eprint = "1111.7241",
    archivePrefix = "arXiv",
    primaryClass = "nucl-ex",
    doi = "10.1016/j.ppnp.2012.01.025",
    journal = "Prog. Part. Nucl. Phys.",
    volume = "67",
    pages = "541--546",
    year = "2012"
}

@article{STAR:2020xiv,
    author = "Adam, Jaroslav and others",
    collaboration = "STAR",
    title = "{Measurement of inclusive charged-particle jet production in Au + Au collisions at $\sqrt{s_{NN}}=$200 GeV}",
    eprint = "2006.00582",
    archivePrefix = "arXiv",
    primaryClass = "nucl-ex",
    doi = "10.1103/PhysRevC.102.054913",
    journal = "Phys. Rev. C",
    volume = "102",
    number = "5",
    pages = "054913",
    year = "2020"
}

@article{CMSRidgepp,
    author = "Khachatryan, Vardan and others",
    collaboration = "CMS",
    title = "{Observation of long-range near-side angular correlations in proton-proton collisions at the LHC}",
    eprint = "1009.4122",
    archivePrefix = "arXiv",
    primaryClass = "hep-ex",
    reportNumber = "CMS-QCD-10-002, CERN-PH-EP-2010-031",
    doi = "10.1007/JHEP09(2010)091",
    journal = "JHEP",
    volume = "09",
    pages = "091",
    year = "2010"
}

@article{Mantysaari:2025tcg,
    author = {M{\"a}ntysaari, Heikki and Schenke, Bj{\"o}rn and Shen, Chun and Zhao, Wenbin},
    title = "{Collision-Energy Dependence in Heavy-Ion Collisions from Nonlinear QCD Evolution}",
    eprint = "2502.05138",
    archivePrefix = "arXiv",
    primaryClass = "nucl-th",
    doi = "10.1103/gf4y-p5j7",
    journal = "Phys. Rev. Lett.",
    volume = "135",
    number = "2",
    pages = "022302",
    year = "2025"
}

@article{nPDFPredictionsPaper,
    author = "Jonas, Florian and Loizides, Constantin and Mazeliauskas, Aleksas and Paakkinen, Petja and Strangmann, Nicolas",
    title = "{A compendium of cold-nuclear matter baseline predictions in light-ion collisions}",
    eprint = "2602.15928",
    archivePrefix = "arXiv",
    primaryClass = "hep-ph",
    month = "2",
    year = "2026"
}

@article{DNNInterpolationPaper,
    author = {Calmon Behling, Maria A. and Kr{\"u}ger, Mario and Jung, Jerome and B{\"u}sching, Henner},
    title = "{DNN predictions for pp reference $p_\mathrm{T}$ spectra at unmeasured $\sqrt{s}$}",
    eprint = "2605.12490",
    archivePrefix = "arXiv",
    primaryClass = "hep-ex",
    month = "5",
    year = "2026"
}

@article{ALICE-PUBLIC-2026-002,
    collaboration = "ALICE",
    author = "Abdallah, Dana Ali Hassan and others",
    title = "{Supplemental material: Evidence for parton energy loss in oxygen-oxygen collisions at $\sqrt{s_{\rm NN}}$ = 5.36 TeV}",
    journal = "{\normalfont\href{https://cds.cern.ch/record/2964308}{ALICE-PUBLIC-2026-002}}",
    year = "2026",
    pages = 17
}

@article{Kudinoor:2026wcs,
    author = "Kudinoor, Arjun Srinivasan and Lin, Arthur Yi-Ting and Pablos, Daniel and Rajagopal, Krishna",
    title = "{A Breath of Fresh Air for Moli{\`e}re: Detecting Moli{\`e}re Scattering using Jet Substructure Observables in Oxygen Collisions}",
    eprint = "2603.23596",
    archivePrefix = "arXiv",
    primaryClass = "hep-ph",
    reportNumber = "MIT-CTP/6018",
    month = "3",
    year = "2026"
}

@article{Mazeliauskas:2025clt,
    author = "Mazeliauskas, Aleksas",
    title = "{Energy loss baseline for light hadrons in oxygen-oxygen collisions at $\sqrt{s_\mathrm{NN}}=5.36\,\text{TeV}$}",
    eprint = "2509.07008",
    archivePrefix = "arXiv",
    primaryClass = "hep-ph",
    doi = "10.1016/j.physletb.2026.140409",
    journal = "Phys. Lett. B",
    volume = "876",
    pages = "140409",
    year = "2026"
}

@article{WA98:2001lnw,
    author = "Aggarwal, M. M. and others",
    collaboration = "WA98",
    title = "{Transverse mass distributions of neutral pions from Pb-208 induced reactions at 158-A-GeV}",
    eprint = "nucl-ex/0108006",
    archivePrefix = "arXiv",
    doi = "10.1007/s100520100886",
    journal = "Eur. Phys. J. C",
    volume = "23",
    pages = "225--236",
    year = "2002"
}

@article{dEnterria:2004cly,
    author = "d'Enterria, David G.",
    title = "{Indications of suppressed high $p_T$ hadron production in nucleus - nucleus collisions at CERN-SPS}",
    eprint = "nucl-ex/0403055",
    archivePrefix = "arXiv",
    doi = "10.1016/j.physletb.2004.06.071",
    journal = "Phys. Lett. B",
    volume = "596",
    pages = "32--43",
    year = "2004"
}

@article{ALICE:2024vzv,
    author = "Acharya, Shreyasi and others",
    collaboration = "ALICE",
    title = "{Observation of partonic flow in proton{\textemdash}proton and proton{\textemdash}nucleus collisions}",
    eprint = "2411.09323",
    archivePrefix = "arXiv",
    primaryClass = "nucl-ex",
    reportNumber = "CERN-EP-2024-299",
    doi = "10.1038/s41467-025-67795-1",
    journal = "Nature Commun.",
    volume = "17",
    number = "1",
    pages = "2585",
    year = "2026"
}

@article{Arleo:2020eia,
    author = "Arleo, Fran{\c{c}}ois and Peign{\'e}, St{\'e}phane",
    title = "{Quenching of Light Hadron Spectra in $p$-A Collisions from Fully Coherent Energy Loss}",
    eprint = "2003.01987",
    archivePrefix = "arXiv",
    primaryClass = "hep-ph",
    doi = "10.1103/PhysRevLett.125.032301",
    journal = "Phys. Rev. Lett.",
    volume = "125",
    number = "3",
    pages = "032301",
    year = "2020"
}

@article{Arleo:2020hat,
    author = "Arleo, Fran{\c{c}}ois and Cougoulic, Florian and Peign{\'e}, St{\'e}phane",
    title = "{Fully coherent energy loss effects on light hadron production in pA collisions}",
    eprint = "2003.06337",
    archivePrefix = "arXiv",
    primaryClass = "hep-ph",
    doi = "10.1007/JHEP09(2020)190",
    journal = "JHEP",
    volume = "09",
    pages = "190",
    year = "2020"
}

@article{Arleo:2021krm,
    author = "Arleo, Fran{\c{c}}ois and Jackson, Greg and Peign{\'e}, St{\'e}phane",
    title = "{Depletion of atmospheric neutrino fluxes from parton energy loss}",
    eprint = "2112.10791",
    archivePrefix = "arXiv",
    primaryClass = "hep-ph",
    reportNumber = "INT-PUB-21-037",
    doi = "10.1016/j.physletb.2022.137541",
    journal = "Phys. Lett. B",
    volume = "835",
    pages = "137541",
    year = "2022"
}

@article{Gebhard:2024flv,
    author = "Gebhard, Jannis and Mazeliauskas, Aleksas and Takacs, Adam",
    title = "{No-quenching baseline for energy loss signals in oxygen-oxygen collisions}",
    eprint = "2410.22405",
    archivePrefix = "arXiv",
    primaryClass = "hep-ph",
    doi = "10.1007/JHEP04(2025)034",
    journal = "JHEP",
    volume = "04",
    pages = "034",
    year = "2025"
}

@article{ALICE:2021leo,
    collaboration = "ALICE",
    author = "Acharya, Shreyasi and others",
    title = "{ALICE 2016-2017-2018 luminosity determination for pp collisions at $\mathbf{\sqrt{{\textit s}}}$ = 13 TeV}",
    journal = "{\normalfont\href{https://cds.cern.ch/record/2776672}{ALICE-PUBLIC-2021-005}}",
    year = "2021",
    pages = 24
}

@article{CMS:2011iwn,
    author = "Chatrchyan, Serguei and others",
    collaboration = "CMS",
    title = "{Observation and studies of jet quenching in PbPb collisions at nucleon-nucleon center-of-mass energy = 2.76 TeV}",
    eprint = "1102.1957",
    archivePrefix = "arXiv",
    primaryClass = "nucl-ex",
    reportNumber = "CERN-PH-EP-2011-001, CMS-HIN-10-004",
    doi = "10.1103/PhysRevC.84.024906",
    journal = "Phys. Rev. C",
    volume = "84",
    pages = "024906",
    year = "2011"
}

@article{ATLAS:2010isq,
    author = "Aad, Georges and others",
    collaboration = "ATLAS",
    title = "{Observation of a Centrality-Dependent Dijet Asymmetry in Lead-Lead Collisions at $\sqrt{s_{NN}}=2.77$ TeV with the ATLAS Detector at the LHC}",
    eprint = "1011.6182",
    archivePrefix = "arXiv",
    primaryClass = "hep-ex",
    doi = "10.1103/PhysRevLett.105.252303",
    journal = "Phys. Rev. Lett.",
    volume = "105",
    pages = "252303",
    year = "2010"
}

\newpage
\appendix

\section{The ALICE Collaboration}
\label{app:collab}
\begin{flushleft} 
\small

D.A.H.~Abdallah\,\orcidlink{0000-0003-4768-2718}\,$^{\rm 134}$, 
I.J.~Abualrob\,\orcidlink{0009-0005-3519-5631}\,$^{\rm 112}$, 
S.~Acharya\,\orcidlink{0000-0002-9213-5329}\,$^{\rm 49}$, 
K.~Agarwal\,\orcidlink{0000-0001-5781-3393}\,$^{\rm II,}$$^{\rm 23}$, 
G.~Aglieri Rinella\,\orcidlink{0000-0002-9611-3696}\,$^{\rm 32}$, 
L.~Aglietta\,\orcidlink{0009-0003-0763-6802}\,$^{\rm 24}$, 
N.~Agrawal\,\orcidlink{0000-0003-0348-9836}\,$^{\rm 25}$, 
Z.~Ahammed\,\orcidlink{0000-0001-5241-7412}\,$^{\rm 132}$, 
S.~Ahmad\,\orcidlink{0000-0003-0497-5705}\,$^{\rm 15}$, 
Z.~Akbar$^{\rm 79}$
% ...

\section*{Affiliation Notes}

$^{\rm I}$ Deceased\\
$^{\rm II}$ Also at: INFN Trieste, Trieste, Italy\\
$^{\rm III}$ Also at: Fondazione Bruno Kessler (FBK), Trento, Italy\\
$^{\rm IV}$ Also at: Czech Technical University in Prague, Prague, Czech Republic\\
$^{\rm V}$ Also at: Instituto de Fisica da Universidade de Sao Paulo\\
$^{\rm VI}$ Also at: Dipartimento DET del Politecnico di Torino, Turin, Italy\\
$^{\rm VII}$ Also at: University College of Dublin, Dublin, Ireland\\
$^{\rm VIII}$ Also at: Department of Applied Physics, Aligarh Muslim University, Aligarh, India\\
$^{\rm IX}$ Also at: Institute of Theoretical Physics, University of Wroclaw, Wroclaw, Poland\\
$^{\rm X}$ Also at: Facultad de Ciencias, Universidad Nacional Aut\'{o}noma de M\'{e}xico, Mexico City, Mexico\\

\section*{Collaboration Institutes}

$^{1}$ A.I. Alikhanyan National Science Laboratory (Yerevan Physics Institute) Foundation, Yerevan, Armenia\\
$^{2}$ AGH University of Krakow, Cracow, Poland\\
$^{3}$ Bogolyubov Institute for Theoretical Physics, National Academy of Sciences of Ukraine, Kyiv, Ukraine\\
$^{4}$ Bose Institute, Department of Physics  and Centre for Astroparticle Physics and Space Science (CAPSS), Kolkata, India\\
$^{5}$ California Polytechnic State University, San Luis Obispo, California, United States\\
$^{6}$ Central China Normal University, Wuhan, China\\
$^{7}$ Centro de Aplicaciones Tecnol\'{o}gicas y Desarrollo Nuclear (CEADEN), Havana, Cuba\\
$^{8}$ Centro de Investigaci\'{o}n y de Estudios Avanzados (CINVESTAV), Mexico City and M\'{e}rida, Mexico\\
$^{9}$ Chicago State University, Chicago, Illinois, United States\\
$^{10}$ China Nuclear Data Center, China Institute of Atomic Energy, Beijing, China\\
$^{11}$ China University of Geosciences, Wuhan, China\\
$^{12}$ Chungbuk National University, Cheongju, Republic of Korea\\
$^{13}$ Comenius University Bratislava, Faculty of Mathematics, Physics and Informatics, Bratislava, Slovak Republic\\
$^{14}$ Creighton University, Omaha, Nebraska, United States\\
$^{15}$ Department of Physics, Aligarh Muslim University, Aligarh, India\\
$^{16}$ Department of Physics, Pusan National University, Pusan, Republic of Korea\\
$^{17}$ Department of Physics, Sejong University, Seoul, Republic of Korea\\
$^{18}$ Department of Physics, University of California, Berkeley, California, United States\\
$^{19}$ Department of Physics, University of Oslo, Oslo, Norway\\
$^{20}$ Department of Physics and Technology, University of Bergen, Bergen, Norway\\
$^{21}$ Dipartimento di Fisica, Universit\`{a} di Pavia, Pavia, Italy\\
$^{22}$ Dipartimento di Fisica dell'Universit\`{a} and Sezione INFN, Cagliari, Italy\\
$^{23}$ Dipartimento di Fisica dell'Universit\`{a} and Sezione INFN, Trieste, Italy\\
$^{24}$ Dipartimento di Fisica dell'Universit\`{a} and Sezione INFN, Turin, Italy\\
$^{25}$ Dipartimento di Fisica e Astronomia dell'Universit\`{a} and Sezione INFN, Bologna, Italy\\
$^{26}$ Dipartimento di Fisica e Astronomia dell'Universit\`{a} and Sezione INFN, Catania, Italy\\
$^{27}$ Dipartimento di Fisica e Astronomia dell'Universit\`{a} and Sezione INFN, Padova, Italy\\
$^{28}$ Dipartimento di Fisica `E.R.~Caianiello' dell'Universit\`{a} and Gruppo Collegato INFN, Salerno, Italy\\
$^{29}$ Dipartimento DISAT del Politecnico and Sezione INFN, Turin, Italy\\
$^{30}$ Dipartimento di Scienze MIFT, Universit\`{a} di Messina, Messina, Italy\\
$^{31}$ Dipartimento Interateneo di Fisica `M.~Merlin' and Sezione INFN, Bari, Italy\\
$^{32}$ European Organization for Nuclear Research (CERN), Geneva, Switzerland\\
$^{33}$ Faculty of Electrical Engineering, Mechanical Engineering and Naval Architecture, University of Split, Split, Croatia\\
$^{34}$ Faculty of Nuclear Sciences and Physical Engineering, Czech Technical University in Prague, Prague, Czech Republic\\
$^{35}$ Faculty of Physics, Sofia University, Sofia, Bulgaria\\
$^{36}$ Faculty of Science, P.J.~\v{S}af\'{a}rik University, Ko\v{s}ice, Slovak Republic\\
$^{37}$ Faculty of Technology, Environmental and Social Sciences, Bergen, Norway\\
$^{38}$ Frankfurt Institute for Advanced Studies, Johann Wolfgang Goethe-Universit\"{a}t Frankfurt, Frankfurt, Germany\\
$^{39}$ Fudan University, Shanghai, China\\
$^{40}$ Gauhati University, Department of Physics, Guwahati, India\\
$^{41}$ Helmholtz-Institut f\"{u}r Strahlen- und Kernphysik, Rheinische Friedrich-Wilhelms-Universit\"{a}t Bonn, Bonn, Germany\\
$^{42}$ Helsinki Institute of Physics (HIP), Helsinki, Finland\\
$^{43}$ High Energy Physics Group,  Universidad Aut\'{o}noma de Puebla, Puebla, Mexico\\
$^{44}$ Horia Hulubei National Institute of Physics and Nuclear Engineering, Bucharest, Romania\\
$^{45}$ HUN-REN Wigner Research Centre for Physics, Budapest, Hungary\\
$^{46}$ Indian Institute of Technology Bombay (IIT), Mumbai, India\\
$^{47}$ Indian Institute of Technology Indore, Indore, India\\
$^{48}$ INFN, Laboratori Nazionali di Frascati, Frascati, Italy\\
$^{49}$ INFN, Sezione di Bari, Bari, Italy\\
$^{50}$ INFN, Sezione di Bologna, Bologna, Italy\\
$^{51}$ INFN, Sezione di Cagliari, Cagliari, Italy\\
$^{52}$ INFN, Sezione di Catania, Catania, Italy\\
$^{53}$ INFN, Sezione di Padova, Padova, Italy\\
$^{54}$ INFN, Sezione di Pavia, Pavia, Italy\\
$^{55}$ INFN, Sezione di Torino, Turin, Italy\\
$^{56}$ INFN, Sezione di Trieste, Trieste, Italy\\
$^{57}$ Inha University, Incheon, Republic of Korea\\
$^{58}$ Institute for Gravitational and Subatomic Physics (GRASP), Utrecht University/Nikhef, Utrecht, Netherlands\\
$^{59}$ Institute of Experimental Physics, Slovak Academy of Sciences, Ko\v{s}ice, Slovak Republic\\
$^{60}$ Institute of Physics, Homi Bhabha National Institute, Bhubaneswar, India\\
$^{61}$ Institute of Physics of the Czech Academy of Sciences, Prague, Czech Republic\\
$^{62}$ Institute of Space Science (ISS), Bucharest, Romania\\
$^{63}$ Institut f\"{u}r Kernphysik, Johann Wolfgang Goethe-Universit\"{a}t Frankfurt, Frankfurt, Germany\\
$^{64}$ Instituto de Ciencias Nucleares, Universidad Nacional Aut\'{o}noma de M\'{e}xico, Mexico City, Mexico\\
$^{65}$ Instituto de F\'{i}sica, Universidade Federal do Rio Grande do Sul (UFRGS), Porto Alegre, Brazil\\
$^{66}$ Instituto de F\'{\i}sica, Universidad Nacional Aut\'{o}noma de M\'{e}xico, Mexico City, Mexico\\
$^{67}$ iThemba LABS, National Research Foundation, Somerset West, South Africa\\
$^{68}$ Jeonbuk National University, Jeonju, Republic of Korea\\
$^{69}$ Korea Institute of Science and Technology Information, Daejeon, Republic of Korea\\
$^{70}$ Laboratoire de Physique Subatomique et de Cosmologie, Universit\'{e} Grenoble-Alpes, CNRS-IN2P3, Grenoble, France\\
$^{71}$ Lawrence Berkeley National Laboratory, Berkeley, California, United States\\
$^{72}$ Lund University Department of Physics, Division of Particle Physics, Lund, Sweden\\
$^{73}$ Marietta Blau Institute, Vienna, Austria\\
$^{74}$ Nagasaki Institute of Applied Science, Nagasaki, Japan\\
$^{75}$ Nara Women{'}s University (NWU), Nara, Japan\\
$^{76}$ National Centre for Nuclear Research, Warsaw, Poland\\
$^{77}$ National Institute of Science Education and Research, Homi Bhabha National Institute, Jatni, India\\
$^{78}$ National Nuclear Research Center, Baku, Azerbaijan\\
$^{79}$ National Research and Innovation Agency - BRIN, Jakarta, Indonesia\\
$^{80}$ Niels Bohr Institute, University of Copenhagen, Copenhagen, Denmark\\
$^{81}$ Nikhef, National institute for subatomic physics, Amsterdam, Netherlands\\
$^{82}$ Nuclear Physics Group, STFC Daresbury Laboratory, Daresbury, United Kingdom\\
$^{83}$ Nuclear Physics Institute of the Czech Academy of Sciences, Husinec-\v{R}e\v{z}, Czech Republic\\
$^{84}$ Oak Ridge National Laboratory, Oak Ridge, Tennessee, United States\\
$^{85}$ Ohio State University, Columbus, Ohio, United States\\
$^{86}$ Physics department, Faculty of science, University of Zagreb, Zagreb, Croatia\\
$^{87}$ Physics Department, Panjab University, Chandigarh, India\\
$^{88}$ Physics Department, University of Jammu, Jammu, India\\
$^{89}$ Physics Program and International Institute for Sustainability with Knotted Chiral Meta Matter (WPI-SKCM$^{2}$), Hiroshima University, Hiroshima, Japan\\
$^{90}$ Physikalisches Institut, Eberhard-Karls-Universit\"{a}t T\"{u}bingen, T\"{u}bingen, Germany\\
$^{91}$ Physikalisches Institut, Ruprecht-Karls-Universit\"{a}t Heidelberg, Heidelberg, Germany\\
$^{92}$ Physik Department, Technische Universit\"{a}t M\"{u}nchen, Munich, Germany\\
$^{93}$ Politecnico di Bari and Sezione INFN, Bari, Italy\\
$^{94}$ Research Division and ExtreMe Matter Institute EMMI, GSI Helmholtzzentrum f\"ur Schwerionenforschung GmbH, Darmstadt, Germany\\
$^{95}$ Saga University, Saga, Japan\\
$^{96}$ Saha Institute of Nuclear Physics, Homi Bhabha National Institute, Kolkata, India\\
$^{97}$ School of Physics and Astronomy, University of Birmingham, Birmingham, United Kingdom\\
$^{98}$ Secci\'{o}n F\'{\i}sica, Departamento de Ciencias, Pontificia Universidad Cat\'{o}lica del Per\'{u}, Lima, Peru\\
$^{99}$ SUBATECH, IMT Atlantique, Nantes Universit\'{e}, CNRS-IN2P3, Nantes, France\\
$^{100}$ Sungkyunkwan University, Suwon City, Republic of Korea\\
$^{101}$ Suranaree University of Technology, Nakhon Ratchasima, Thailand\\
$^{102}$ Technical University of Ko\v{s}ice, Ko\v{s}ice, Slovak Republic\\
$^{103}$ The Henryk Niewodniczanski Institute of Nuclear Physics, Polish Academy of Sciences, Cracow, Poland\\
$^{104}$ The University of Texas at Austin, Austin, Texas, United States\\
$^{105}$ Universidad Aut\'{o}noma de Sinaloa, Culiac\'{a}n, Mexico\\
$^{106}$ Universidade de S\~{a}o Paulo (USP), S\~{a}o Paulo, Brazil\\
$^{107}$ Universidade Estadual de Campinas (UNICAMP), Campinas, Brazil\\
$^{108}$ Universidade Federal do ABC, Santo Andre, Brazil\\
$^{109}$ Universitatea Nationala de Stiinta si Tehnologie Politehnica Bucuresti, Bucharest, Romania\\
$^{110}$ University of Cape Town, Cape Town, South Africa\\
$^{111}$ University of Derby, Derby, United Kingdom\\
$^{112}$ University of Houston, Houston, Texas, United States\\
$^{113}$ University of Jyv\"{a}skyl\"{a}, Jyv\"{a}skyl\"{a}, Finland\\
$^{114}$ University of Kansas, Lawrence, Kansas, United States\\
$^{115}$ University of Liverpool, Liverpool, United Kingdom\\
$^{116}$ University of Science and Technology of China, Hefei, China\\
$^{117}$ University of Silesia in Katowice, Katowice, Poland\\
$^{118}$ University of South-Eastern Norway, Kongsberg, Norway\\
$^{119}$ University of Tennessee, Knoxville, Tennessee, United States\\
$^{120}$ University of the Witwatersrand, Johannesburg, South Africa\\
$^{121}$ University of Tokyo, Tokyo, Japan\\
$^{122}$ University of Tsukuba, Tsukuba, Japan\\
$^{123}$ Universit\"{a}t M\"{u}nster, Institut f\"{u}r Kernphysik, M\"{u}nster, Germany\\
$^{124}$ Universit\'{e} Clermont Auvergne, CNRS/IN2P3, LPC, Clermont-Ferrand, France\\
$^{125}$ Universit\'{e} de Lyon, CNRS/IN2P3, Institut de Physique des 2 Infinis de Lyon, Lyon, France\\
$^{126}$ Universit\'{e} de Strasbourg, CNRS, IPHC UMR 7178, F-67000 Strasbourg, France, Strasbourg, France\\
$^{127}$ Universit\'{e} Paris-Saclay, Centre d'Etudes de Saclay (CEA), IRFU, D\'{e}partment de Physique Nucl\'{e}aire (DPhN), Saclay, France\\
$^{128}$ Universit\'{e}  Paris-Saclay, CNRS/IN2P3, IJCLab, Orsay, France\\
$^{129}$ Universit\`{a} degli Studi di Foggia, Foggia, Italy\\
$^{130}$ Universit\`{a} del Piemonte Orientale, Vercelli, Italy\\
$^{131}$ Universit\`{a} di Brescia, Brescia, Italy\\
$^{132}$ Variable Energy Cyclotron Centre, Homi Bhabha National Institute, Kolkata, India\\
$^{133}$ Warsaw University of Technology, Warsaw, Poland\\
$^{134}$ Wayne State University, Detroit, Michigan, United States\\
$^{135}$ Yale University, New Haven, Connecticut, United States\\
$^{136}$ Yildiz Technical University, Istanbul, Turkey\\
$^{137}$ Yonsei University, Seoul, Republic of Korea\\
$^{138}$ Affiliated with an institute formerly covered by a cooperation agreement with CERN\\
$^{139}$ Affiliated with an international laboratory covered by a cooperation agreement with CERN.\\

\end{flushleft} 

\end{document}